\documentclass{iopjournal}

\usepackage[backend=biber, style=chem-angew]{biblatex}

\begin{document}


\vspace*{1mm}

\title{Observation of resonant doublet and variable finesse in a tabletop meter--scale 
linear three--mirror cavity}

\author{Paul Stevens$^1$\orcid{0009-0008-1382-2990},
Pierre--Emmanuel Bonningues$^1$\orcid{0009-0007-5957-9706},
Th\'eo Lesieur$^1$\orcid{0009-0000-1146-7915}
François Glotin$^{1,*}$\orcid{0000-0003-2637-1187},
Vincent Loriette$^1$\orcid{0000-0002-4551-2902},
Manuel Andia$^1$\orcid{0000-0003-3675-9126},
Angélique Lartaux--Vollard$^1$\orcid{0000-0003-1714-365X},
Nicolas Leroy$^1$\orcid{0000-0002-2321-1017},
and Aymeric van de Walle$^1$\orcid{0000-0003-4573-1109}}

\affil{$^1$Université Paris--Saclay, CNRS/IN2P3, IJCLab, 91405 Orsay, France}

\affil{$^*$Author to whom any correspondence should be addressed.}

\email{francois.glotin@ijclab.in2p3.fr}


\begin{abstract}
Fabry--Perot cavities are widely used in current gravitational--wave detectors. In particular, they play a key role in frequency--dependent squeezing systems, enabling broadband quantum noise reduction. However, their ability to precisely control squeezing properties may be insufficient for the next generation of detectors. In this context, theoretical studies on linear three--mirror cavities have revealed promising features, such as resonance peak splitting and their equivalence to two--mirror cavities with variable finesse. In this paper, we report experimental observations of both the resonant doublet and variable finesse using a meter--scale implementation of a linear three--mirror cavity.
\end{abstract}

\section{Introduction}

Despite the remarkable technological achievements that enable gravitational--wave detectors (GWDs) to measure sub--femtometric space--time distortions, many research and development efforts are still underway to reduce the noise sources that continue to limit their sensitivity. Among these, quantum noise is one of the most significant. In current--generation GWDs \cite{ligo, virgo, kagra}, broadband quantum noise reduction is achieved using squeezed states of light \cite{acernese_increasing_2019, tse_quantum-enhanced_2019}, whose ellipse angle is made frequency--dependent via a two--mirror Fabry--Perot filter cavity \cite{ganapathy_broadband_2023}.

In contrast, proposed next--generation detectors \cite{et_steering_committee_et_2020, cosmicExplorer} will require more sophisticated control of squeezing owing to their more advanced design and operational configuration, necessitating filtering systems with greater flexibility than conventional two--mirror cavities. Among the alternatives under consideration, linear three--mirror cavities (3MCs) appear particularly promising, thanks to their unique optical characteristics. Specifically, this configuration shows two key features of interest. First, when the two sub--cavities are tuned close to their respective resonance conditions, the transmission peak splits into a doublet \cite{stadt, hogeveen, thuring, stevens_resonant_2024}. Second, a 3MC is functionally equivalent to a two--mirror cavity with variable finesse if one sub--cavity is treated as a virtual mirror whose effective reflectivity can be changed by adjusting its tuning \cite{strain_experimental_1994, croquette_recent_2023, acernese_variable_2006}. Regarding the application of these optical properties to next--generation detectors such as Einstein Telescope (ET), the doublet structure enables a single 3MC to reproduce the effect of the two sequential two--mirror filter cavities currently planned to allowing two frequency--dependent rotations of the squeezing ellipse within a single device \cite{ding_performance_2025}. On the other hand, as the detector configuration evolves over its lifetime, the quantum noise spectrum may change. Variable--finesse operation provides a practical way to adjust the filter cavity bandwidth accordingly, without requiring hardware modifications such as mirror replacement. This second approach is applicable to ET and, more generally, to other GWDs employing frequency--dependent squeezing for quantum noise reduction, such as Cosmic Explorer, improving both convenience and sustainability.

In this paper, we present the experimental methods used to observe resonant doublet and variable finesse, along with the results obtained from a tabletop, meter--scale implementation of a linear 3MC.

\newpage

\section{Optical characteristics of linear three--mirror cavities}

\begin{figure}[h!]
\centering
        \includegraphics[width=0.43\textwidth]{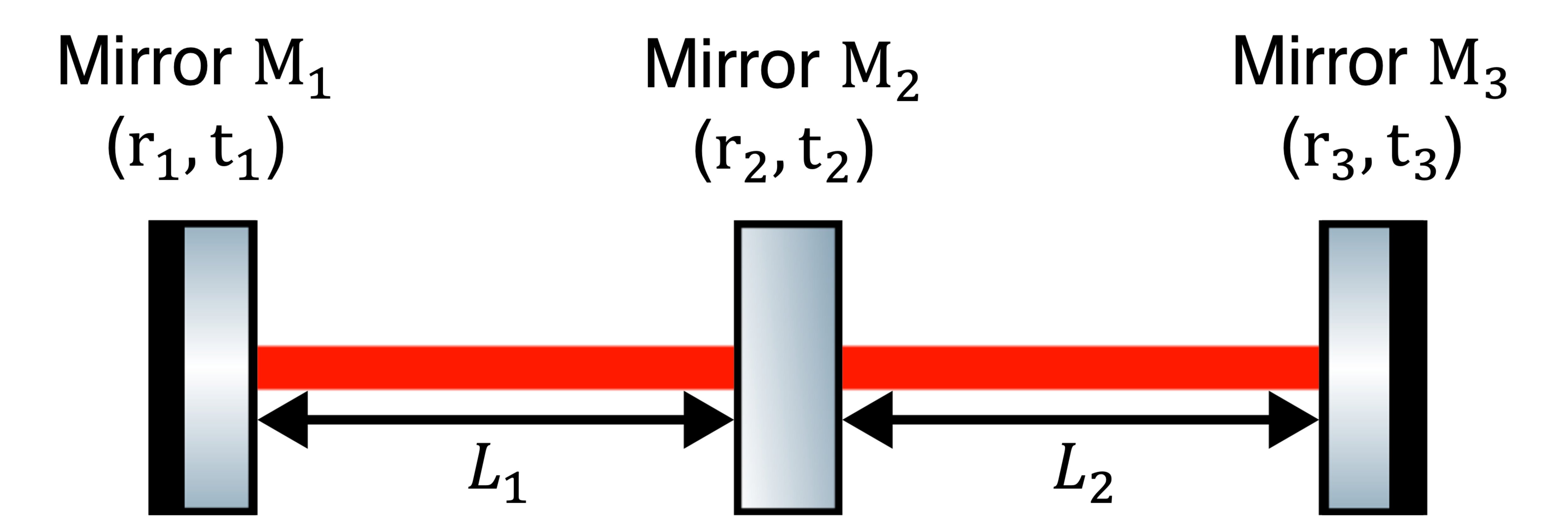}
        \caption{Scheme of a three--mirror cavity.}
\label{fig:figure1}
\end{figure}

A linear 3MC, as schematized in Fig.~\ref{fig:figure1}, consists of three mirrors arranged in series, with the input and end mirrors facing each other as in a standard two--mirror Fabry--Perot cavity. The central mirror is positioned between them and can be oriented towards either direction.

\subsection{Doubling of resonant frequency}

\begin{figure}[h!]
\centering
        \includegraphics[width=0.68\textwidth]{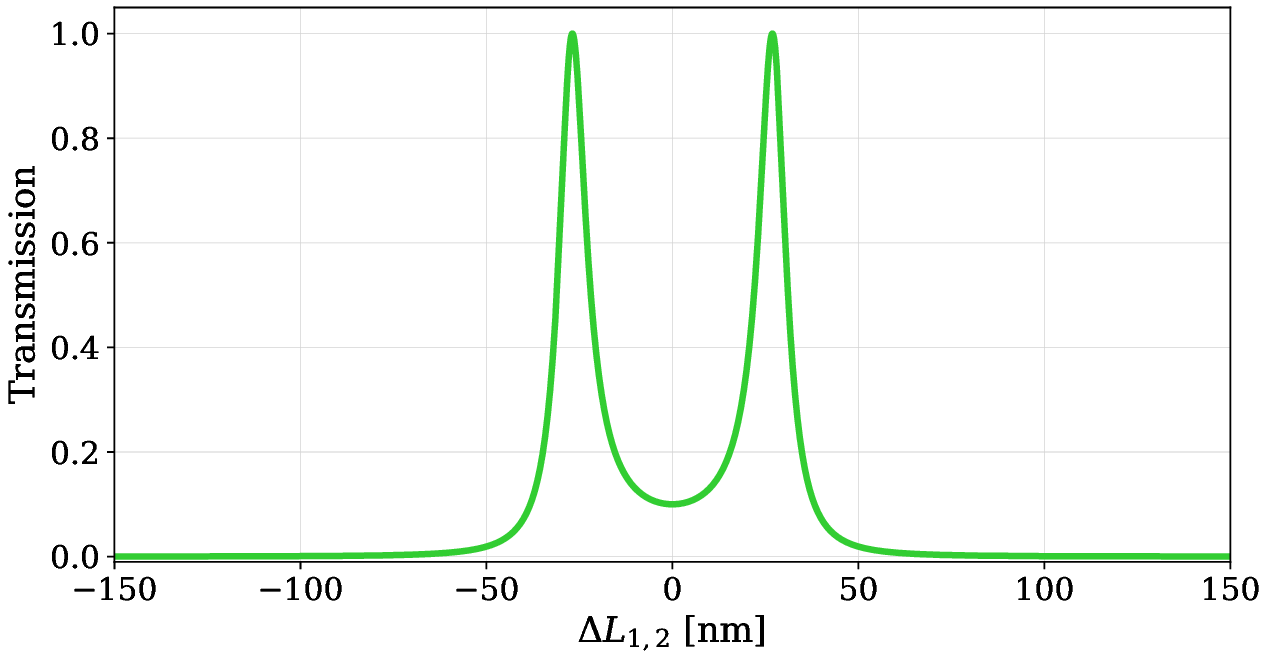}
        \caption{Simulation of the transmission through a three--mirror cavity as a function of the common detuning of the two sub--cavity lengths. The condition $\Delta L_{1,2}=0$ corresponds to resonance in each sub--cavity. Simulation parameters: $L_1=L_2=\SI{0.5}{m}$, $R_1=R_2=R_3=0.9$ and $\lambda=\SI{1064}{nm}$, assuming lossless mirrors.}
\label{fig:figure2}
\end{figure}

The optical response of the 3MC can be characterized by its overall transmission coefficient \cite{stevens_resonant_2024}:
\begin{equation}
    t=\frac{-t_1t_2t_3e^{ik(L_1+L_2)}}{e^{2ik(L_1+L_2)}-r_1r_2e^{2ikL_2}-r_2r_3e^{2ikL_1}+r_1r_3(r_2^2+t_2^2)}
    \label{eq_EMC_t}
\end{equation}
where $k$ is the wave vector of the light source, and the remaining parameters correspond to those defined in Fig.~\ref{fig:figure1}. The multiple interactions and coupling between the sub--cavities are shown in the expression of the transmission coefficient $t$, where the individual contributions of each sub--cavity are no longer distinguishable. This more complex behavior, compared to that of a standard two--mirror cavity, gives rise to a distinctive property of the 3MC: when both sub--cavities are simultaneously near resonance, the single resonant frequency expected from a two--mirror cavity splits into two distinct peaks \cite{stadt, thuring, stevens_resonant_2024}. This results in a characteristic ``double--peak'' resonance pattern, illustrated in Fig.~\ref{fig:figure2}, which can be shaped depending on the cavity configuration.

\subsection{Variable finesse}

\begin{figure}[t!]
\centering
        \includegraphics[width=0.5\textwidth]{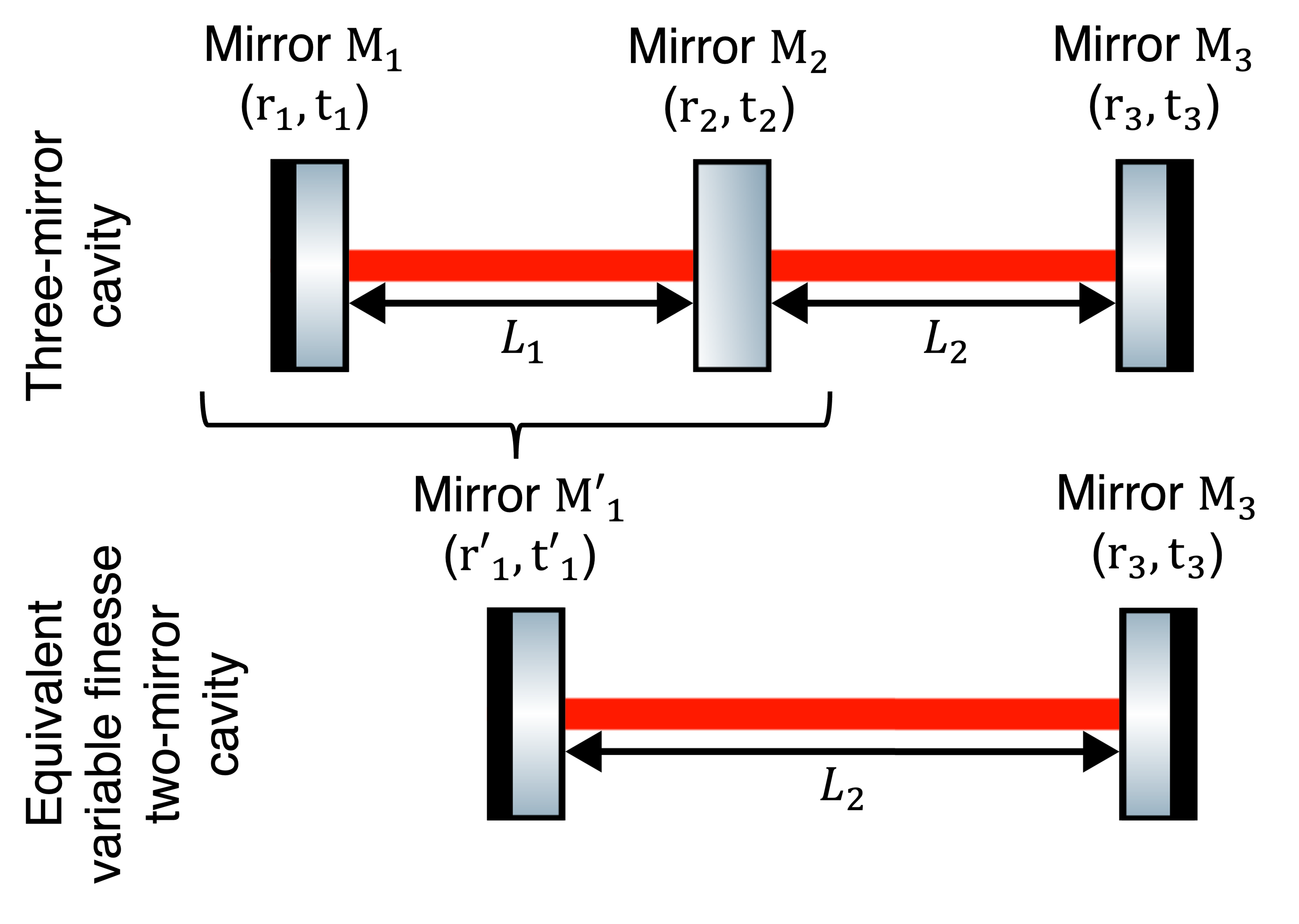}
        \caption{Equivalence between a three--mirror cavity (top) and a virtual two--mirror cavity with variable finesse (bottom). In the three--mirror configuration, mirrors $\mathrm{M}_1$ and $\mathrm{M}_2$ form a sub--cavity parameterized by $(r_1, t_1)$, $(r_2, t_2)$, and length $L_1$. This sub--cavity can be modeled as a virtual mirror $\mathrm{M'}_1$ with tunable reflection and transmission coefficients $(r'_1, t'_1)$. Both configurations share the length $L_2$ and mirror parameters $(r_3, t_3)$.}
\label{fig:figure3}
\end{figure}

Another key feature of the 3MC is its equivalence to a virtual two--mirror Fabry--Perot cavity with variable finesse. As illustrated in Fig.~\ref{fig:figure3}, 
the first sub--cavity in the three--mirror system can be interpreted as a virtual mirror, whose effective reflection and transmission coefficients vary with its tuning \cite{croquette_recent_2023}. In other words, the optical properties of the virtual mirror $\text{M}'_1$ can be tuned by making microscopic adjustments to the length $L_1$. 

Either the first or the second sub--cavity can be chosen to serve as a virtual mirror, however, this choice remains significant for two reasons. First, when the three mirrors have different reflectivity coefficients, it affects the virtual mirror’s reflectivity range and, consequently, the achievable finesse interval. Second, even if all mirrors share the same reflectivity coefficient, the selection remains significant, as it influences the ratio between the input and output mirror reflectivities in the virtual two--mirror cavity, resulting in under--coupled, optimally--coupled, or over--coupled states. 

The standard definition of finesse allows one to express that of the 3MC. In the case where the first sub--cavity is treated as a virtual mirror, the corresponding finesse $\mathcal{F}^{\mathrm{SC1}}_\mathrm{3MC}$ can be expressed as a function of the reflectivities of the virtual mirror $\mathrm{M'}_1$ and the third mirror $\mathrm{M}_3$:
\begin{equation}
    \mathcal{F}^{\mathrm{SC1}}_\mathrm{3MC} = \pi \left[\arccos\left( - \frac{1 + {r'_1}^2 r_3^2 - 4 r'_1 r_3}{2 r'_1 r_3} \right)\right]^{-1}
    \label{eq:3MCfinesse}
\end{equation}
where the effective reflectivity $r'_1$ of $\mathrm{M'}_1$ is given by:
\begin{equation}
    r'_1 = \frac{r_1 - r_2 (r_1^2 + t_1^2) e^{-2ikL_1}}{1 - r_1 r_2 e^{-2ikL_1}}
    \label{eq:r1prime_3MCfinesse}
\end{equation}
Conversely, when the second sub--cavity is treated as a virtual mirror, the finesse of the 3MC can be written in the same form as Eq.~(\ref{eq:3MCfinesse}), yielding $\mathcal{F}^{\mathrm{SC2}}_\mathrm{3MC}$ as a function of $r_1$ and the effective reflectivity $r'_2$ of the virtual mirror $\mathrm{M'}_2$. The latter depends on $r_2$, $r_3$, and $L_2$ in the same manner as Eq.~(\ref{eq:r1prime_3MCfinesse}).

\section{Experimental three--mirror cavity setup}

\begin{figure}[b!]
\centering
        \includegraphics[width=0.5\textwidth]{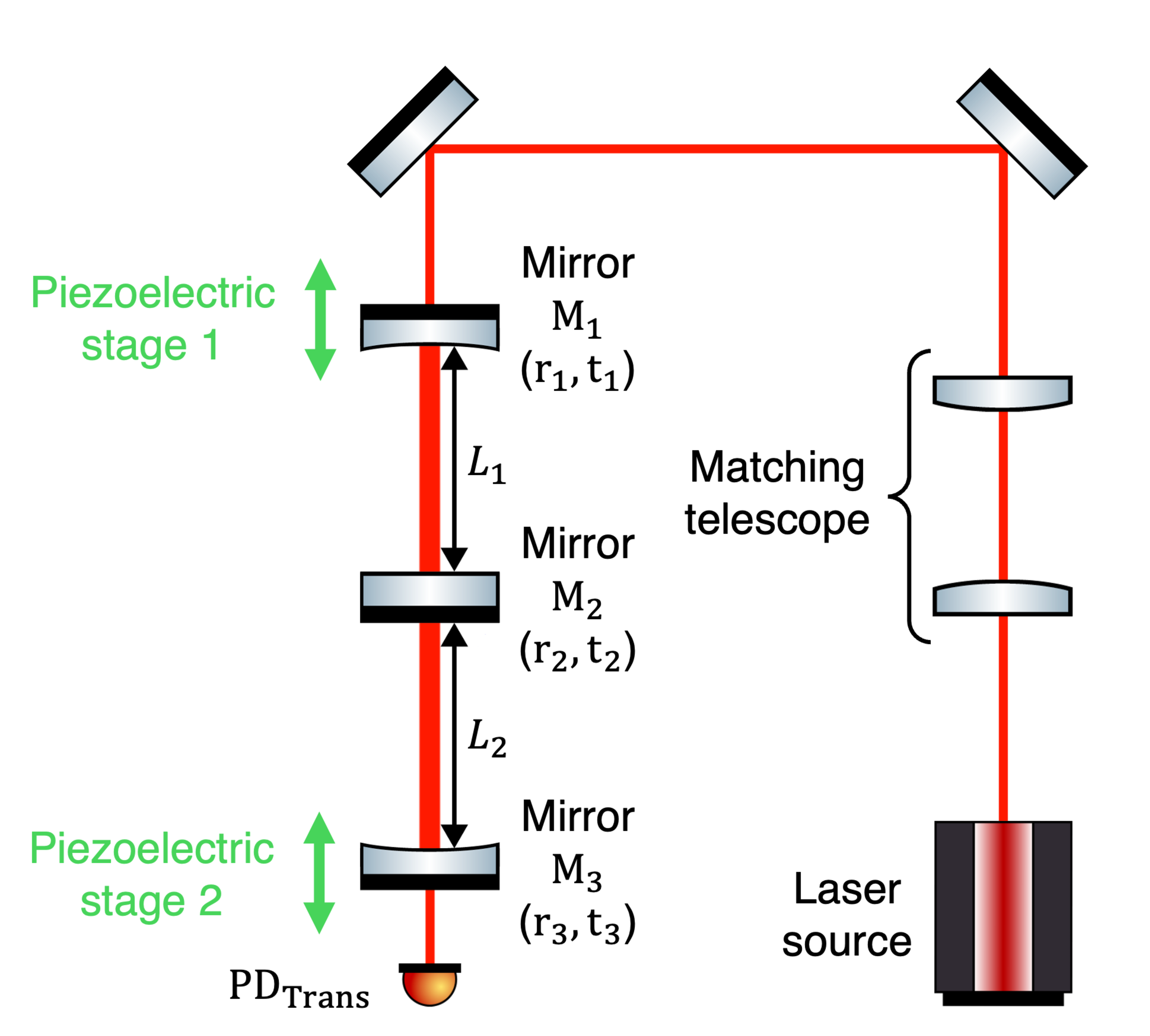}
        \caption{Experimental setup of a three--mirror cavity, constituted by an input mirror $\text{M}_1$, a middle mirror $\text{M}_2$, and an end mirror $\text{M}_3$. Each mirror has amplitude reflection and transmission coefficients $r_i$ and $t_i$, respectively, where $i=1,2,3$. The light source is a \SI{1064}{nm} laser, and its beam profile is matched to the cavity geometry using a two--lens telescope. The transmitted field is monitored by the photodiode $\text{PD}_{\text{Trans}}$. The two outer mirrors, $\text{M}_1$ and $\text{M}_3$, are mounted on piezoelectric stages, both driven by a synchronized signal from the same reference generator.}
\label{fig:figure4}
\end{figure}
Fig.~\ref{fig:figure4} presents the schematic of our experimental setup and the labeling conventions used in the following sections. We implemented a fully symmetric 3MC with a total length of \SI{1}{m}, divided into two equal--length sub--cavities: $L_1=L_2=\SI{0.5}{m}$. All three mirrors from \textit{Laseroptiks} are specified to have the same reflectivity of $R_1=R_2=R_3= \SI{90(2)}{\%}$. To ensure cavity stability, we adopted a convex–plano–convex geometry \cite{stevens_resonant_2024}, where the outer mirrors have a radius of curvature $\rho_1=\rho_3=\SI{1}{m}$. To enable independent scanning of the sub--cavity lengths, the first and third mirrors are mounted on piezoelectric (PZT) translation stages \textit{Piezoconcept HS1}, each with an absolute displacement range of \SI{10}{\micro\metre}, while the central mirror remains fixed. We measured the displacement coefficient of each PZT stage and accordingly calibrated the applied voltage. Finally, the light source is a \textit{Coherent Mephisto S} laser with a wavelength of \SI{1064}{nm}.

\section{Experimental observation of the resonance frequency doubling}

\subsection{Method}

\label{sec:doubletMethod}

\begin{figure}[!h]
\centering
        \includegraphics[width=0.68\textwidth]{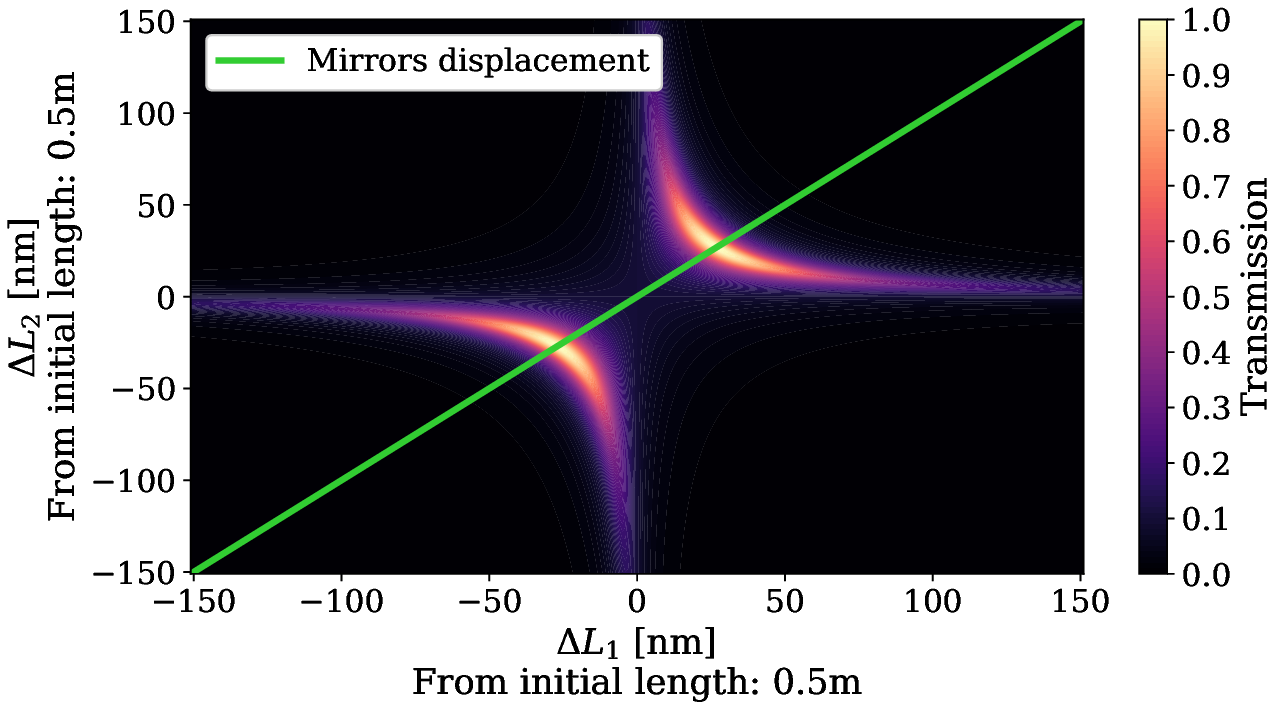}
        \caption{Transmission as a function of the detuning of each sub--cavity. The initial condition $\Delta L_1=\Delta L_2=0$ corresponds to both sub--cavities being individually resonant. The solid green line indicates the sequence of three--mirror cavity states obtained by scanning the positions of $\mathrm{M}_1$ and $\mathrm{M}_3$ using piezoelectric stages moving synchronously in opposite directions with equal amplitude. Simulation parameters: $R_1=R_2=R_3=0.9$ and $\lambda=\SI{1064}{nm}$, assuming lossless mirrors.
        }
\label{fig:figure5}
\end{figure}

Fig.~\ref{fig:figure5} shows a simulation of transmission through our 3MC setup as a function of the detuning of each sub--cavity from its respective resonance condition. The resulting map displays a cross--like pattern, with two dominant resonance branches corresponding to the displacement of the maxima of the resonant doublet. Specifically, the lower--left and upper--right branches represent, respectively, the evolution of the left and right maxima of the doublet shown in Fig.~\ref{fig:figure2}. 

To access the doublet structure experimentally, identical displacement ramps are applied to the two PZT stages controlling the sub--cavity lengths, ensuring synchronous variations of equal amplitude. The 3MC is thus scanned along the diagonal direction indicated by the solid green line in Fig.~\ref{fig:figure5}. Owing to the fully symmetric configuration of the 3MC, with identical mirror reflectivities and equal macroscopic sub--cavity lengths, the doublet is expected to be perfectly symmetric as shown in Fig.~\ref{fig:figure2}, with equal peak width, amplitude, and spacing around the resonance condition in each sub--cavity \cite{stevens_resonant_2024}.

\subsection{Results}

\begin{figure}[!h]
\centering
        \includegraphics[width=0.68\textwidth]{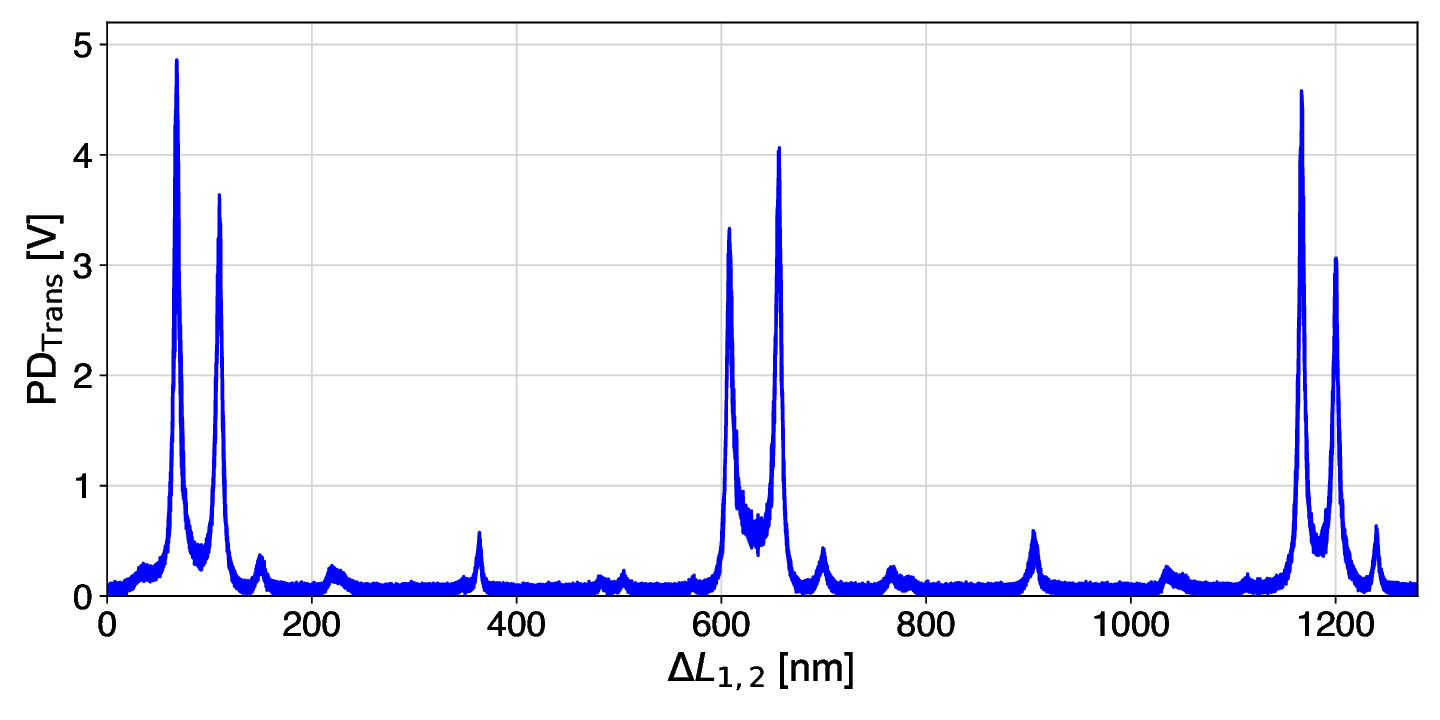}
        \caption{Observation of resonant doublets. Voltage signal from the photodiode $\text{PD}_{\text{Trans}}$ as a function of sub--cavities' common detuning.}
\label{fig:figure6}
\end{figure}

A distinct splitting of the resonant peaks is observed in the experimental data, presented in Fig.~\ref{fig:figure6}. As discussed in section~\ref{sec:doubletMethod}, the resonant doublets expected from the symmetric configuration of the setup are themselves symmetric. In practice, however, the measured transmission spectrum shows asymmetries in both the amplitude ratio and the spacing between the two maxima of each doublet. These irregularities do not follow a systematic trend but instead evolve randomly. This variability can be primarily attributed to two factors. First, slight deviations in mirror reflectivity may be present, and even minimal differences can lead to unequal optical coupling between the two sub--cavities, thereby disrupting the balance between the peaks. A second contributing factor involves the mechanical stability of the central mirror, which, although nominally fixed, is not actively stabilized. It is therefore subject to positional drifts over time, likely induced by external mechanical perturbations such as air currents, vibrations transmitted through the optical bench, and thermal expansion or contraction due to temperature variations. Given that the amplitude ratio of the resonant peaks is highly sensitive to both mirror positioning and reflectivity values \cite{stevens_resonant_2024}, even minor mechanical displacements or small deviations in reflectivity are sufficient to make the measurement particularly challenging, resulting in the observed distortions of the resonance pattern.

Using the calibrated displacement coefficient of the PZT stage, the measured intra--doublet spacing is \SI{42(10)}{\nano\meter}. This value is in good agreement with the theoretically expected spacing of \SI{54(6)}{\nano\meter}, calculated from the nominal mirror reflectivity ${R_\text{th} = \SI{90(2)}{\percent}}$. Furthermore, the separation between neighboring doublets, defined as the distance between the centers of adjacent doublets, is measured to be \SI{555(44)}{\nano\meter}. This result is consistent with the theoretically predicted value of \SI{532}{\nano\meter}, corresponding to the half--wavelength resonance spacing of a laser operating at \SI{1064}{\nano\meter}. For both the intra--doublet spacing and the separation between successive doublets, the uncertainties are evaluated by accounting for the uncertainty in the voltage--to--displacement coefficient of the PZT stages.

In addition to the primary resonance doublets, higher--order resonances are observed between the main peaks. These secondary resonances are attributed to imperfect mode matching between the incident laser beam and the cavity geometry, as well as to slight angular misalignments and positional offsets of the cavity mirrors.

Finally, it is worth noting that the observed sensitivity of the double--peak structure is not expected to constitute a limitation for ET, as discussed in \cite{stevens_study_2025}, where advanced vibration isolation and vacuum systems, as well as stringent requirements on mirror quality will strongly suppress the disturbances affecting the present setup.

\section{Experimental observation of finesse variation}

\subsection{Method}

The variation of the 3MC finesse can be experimentally measured by studying its dependence on the effective reflectivity of the sub--cavity acting as a virtual mirror. The method consists of two steps. First, one sub--cavity is held at a fixed length, thereby defining the effective reflectivity of the corresponding virtual mirror, while the length of the second sub--cavity is scanned. The resulting transmission resonance peaks are then analyzed to extract the associated 3MC finesse. Second, this procedure is repeated for different tunings of the virtual--mirror sub--cavity, enabling the finesse to be mapped as a function of the effective reflectivities of the virtual mirror.

When the sub--cavity acting as the virtual mirror is not actively controlled, as in the present experimental setup, an alternative strategy, referred to as the dual--scan method, can be used. In this approach, the virtual--mirror sub--cavity is scanned at a relatively slow rate, leading to a gradual change in its effective reflectivity, while the second sub--cavity is scanned at a higher frequency. This dual--scan protocol has two main advantages. First, it allows a continuous dataset to be recorded, enabling the finesse to be evaluated throughout all stages of the virtual--mirror reflectivity evolution. Second, it reduces the impact of environmental noise and long--term mechanical drifts, since measurements can be acquired over a relatively short time.

In this study, the dual--scan method is implemented by scanning the first sub--cavity at a frequency of \SI{10}{mHz} and the second sub--cavity at \SI{70}{Hz}. This choice provides a sufficiently large separation of timescales, enabling an accurate determination of the finesse as a function of the detuning of $L_1$, while maintaining a total acquisition time on the order of a few minutes. This relatively short total measurement time further minimizes the impact of thermal drifts in the optical bench associated with temperature fluctuations.

During the measurements, the displacements of mirrors $M_{1}$ and $M_{3}$ are recorded using the PZT displacement sensors integrated into the PZT stages, while the optical transmission of the 3MC is simultaneously recorded via $\text{PD}_{\text{Trans}}$. In the transmission dataset, each pair of adjacent transmission peaks is fitted with an Airy function. To ensure robustness and reliability of the analysis, a quality filter is applied to all peak--pair fits, based on both the reduced $\chi^2$ value of each fit and the uncertainty associated with the extracted finesse, allowing poorly constrained or unreliable fits to be excluded. From the retained fits, the mean FWHM and the FSR are determined, enabling the calculation of the 3MC finesse for the corresponding peak--pair. This procedure is repeated for all pairs across the entire dataset in order to reconstruct the finesse evolution.

\subsection{Results}

\begin{figure}[!h]
\centering
        \includegraphics[width=0.68\textwidth]{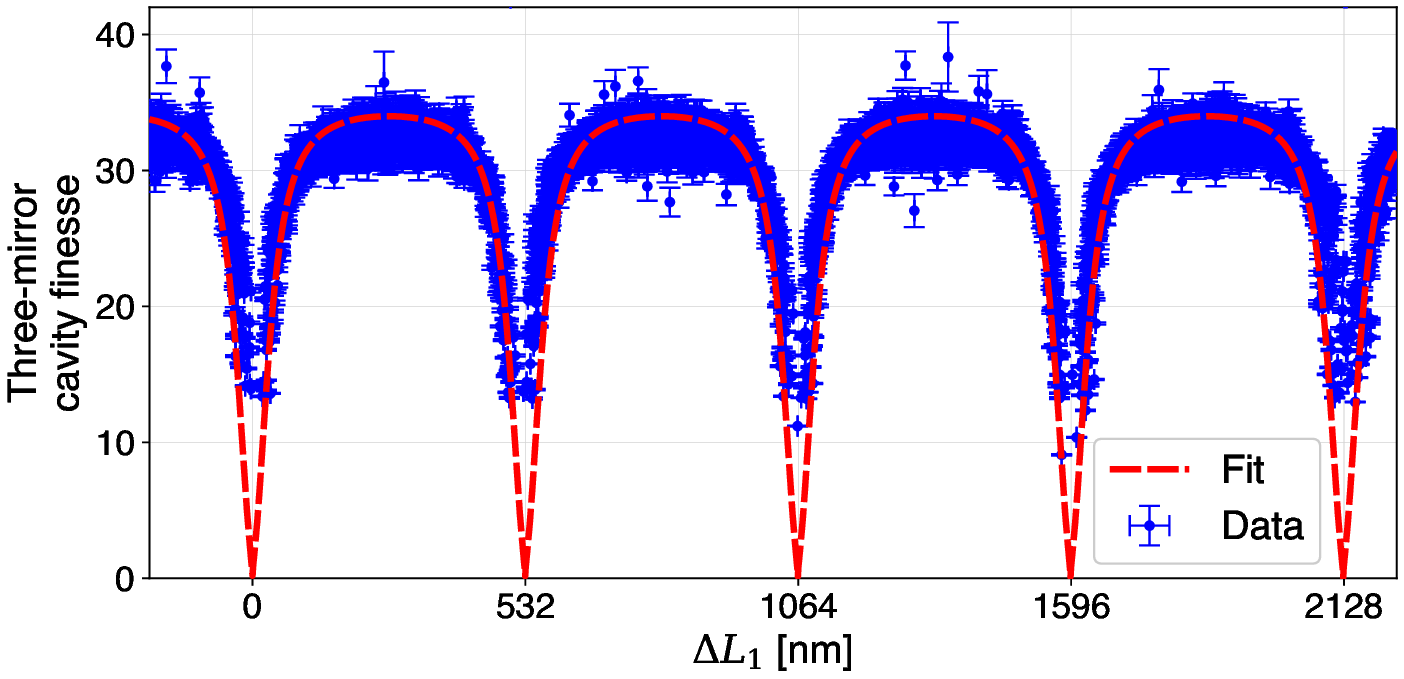}
        \caption{Variable finesse as a function of the detuning $\Delta L_1$ of the first sub--cavity. The finesse is fitted using Eq.~(\ref{eq:3MCfinesse}) and Eq.~(\ref{eq:r1prime_3MCfinesse}). The origin of the x--axis has been shifted to the first finesse drop for better clarity.}
\label{fig:figure7}
\end{figure}

The results of the variable--finesse analysis are presented in Fig.~\ref{fig:figure7}. The evolution of the finesse is characterized by fitting the data to the theoretical model described in Eq.~(\ref{eq:3MCfinesse}) and Eq.~(\ref{eq:r1prime_3MCfinesse}), yielding a finesse modulation periodicity of ${\Delta \mathcal{F}_0 = \SI{532(35)}{\nano\meter}}$. The quoted uncertainty is obtained by propagating the calibration uncertainty of the PZT voltage--to--displacement conversion. These results are consistent with theoretical expectations, showing a drop in finesse at each half--wavelength detuning of the sub--cavity treated as a virtual mirror.

As shown in Fig.~\ref{fig:figure7}, only a limited number of data points are obtained in regions where the finesse approaches zero. This is primarily due to instabilities arising from the mechanical sensitivity of the experimental setup to external perturbations. As the finesse decreases, the FWHM of the resonance peaks increases significantly, theoretically diverging to infinity. Consequently, measurements in this regime become increasingly sensitive, leading to poorly defined resonance peaks and making reliable fitting of the corresponding peak pairs challenging. This limitation is not expected to impact ET, which will employ active mirror control systems to suppress such instabilities and is not designed to operate in regimes of vanishing finesse, in accordance with current design requirements \cite{g_flexibility_2025}.

Finally, Fig.~\ref{fig:figure7} shows that the apparent upper bound of the measured finesse is lower than the maximum theoretical value, $\mathcal{F}^\mathrm{Th, Max}_{\mathrm{3MC}} \approx 58.1^{+15.7}_{-10.5}$, as obtained from Eq.~(\ref{eq:3MCfinesse}) and Eq.~(\ref{eq:r1prime_3MCfinesse}), with uncertainties arising from the propagation of those associated with the mirror reflectivity. This discrepancy is due to the limited precision of the cavity alignment provided by the mirror mounts, which induces coupling from $\text{TEM}_{00}$ into higher--order modes. This leads to an apparent reduction in the effective reflectivity of the virtual mirror as seen from the complementary sub--cavity. This effect is not expected to impact ET, as discussed in \cite{stevens_study_2025}, which will employ active mirror control systems.

\section{Conclusion}

Building upon established theoretical studies of 3MCs, we have successfully designed and implemented a meter--scale experimental prototype, enabling direct observation of resonance peak splitting and variable--finesse behavior. 

Our measurements confirmed the expected resonance peak splitting, which shows good agreement with simulation predictions. Although the simulations are based on idealized assumptions, such as the plane--wave approximation and perfectly lossless mirrors, the close match between theoretical and experimental results validates the robustness of the model.

Furthermore, our results demonstrate that the 3MC can be effectively described as a two--mirror system with a finesse that is variable through detuning of one sub--cavity from resonance. This variability provides a high degree of flexibility, enabling the cavity response to be adapted to different requirements.

In this study, we focused on a fully symmetric configuration, with all three mirrors having identical reflectivities and both sub--cavities set to equal macroscopic lengths. Future investigations will aim to explore varied configurations. This includes asymmetric setups, featuring unequal sub--cavity lengths or differing mirror reflectivities, as well as larger--scale implementations. Additionally, employing suspended mirrors or operating the cavity under vacuum would reduce environmental noise, enabling more precise measurements. Another challenge lies in developing techniques to control such system and lock it in a precise operating regime. Such studies could contribute to the development of filtering systems for the reduction of quantum noise in advanced interferometric GWDs.

\funding{The authors acknowledge the support of the French Agence Nationale de la Recherche (ANR), under Grant Nos. ANR-18-JSTQ-0002 (project QFilter).}

\acknowledgments{We thank the LaseriX and DeLLight teams for their material support, which made these measurements possible. We also thank the LVK collaboration for the fruitful discussions.} 

\printbibliography

@phdthesis{thuring,
	title = {Investigations of coupled and {Kerr} non-linear optical resonators},
	copyright = {Es gilt deutsches Urheberrecht. Das Dokument darf zum eigenen Gebrauch kostenfrei genutzt, aber nicht im Internet bereitgestellt oder an Außenstehende weitergegeben werden.},
	doi = {10.15488/7324},
	language = {en},
	author = {Thüring, André},
	collaborator = {{Technische Informationsbibliothek (TIB)} and {Technische Informationsbibliothek (TIB)}},
	year = {2009},
}

@article{virgo,
	title = {Advanced Virgo: a second-generation interferometric gravitational wave detector},
	volume = {32},
	issn = {0264-9381, 1361-6382},
	doi = {10.1088/0264-9381/32/2/024001},
	shorttitle = {Advanced Virgo},
	pages = {024001},
	number = {2},
	journaltitle = {Classical and Quantum Gravity},
	shortjournal = {Class. Quantum Grav.},
	author = {Acernese, F. and et al.},
	date = {2015-01-22},
}

@article{kagra,
	title = {Detector configuration of {KAGRA}–the Japanese cryogenic gravitational-wave detector},
	volume = {29},
	issn = {0264-9381, 1361-6382},
	doi = {10.1088/0264-9381/29/12/124007},
	pages = {124007},
	number = {12},
	journaltitle = {Classical and Quantum Gravity},
	shortjournal = {Class. Quantum Grav.},
	author = {Somiya, Kentaro and {for the KAGRA Collaboration}},
	date = {2012-06-21},
}

@online{cosmicExplorer,
	title = {Cosmic Explorer: a next generation gravitational wave
    detector},
	url = {https://cosmicexplorer.org},
}

@article{ligo,
	title = {Advanced {LIGO}},
	volume = {32},
	issn = {0264-9381, 1361-6382},
	doi = {10.1088/0264-9381/32/7/074001},
	pages = {074001},
	number = {7},
	journaltitle = {Classical and Quantum Gravity},
	shortjournal = {Class. Quantum Grav.},
	author = {{The LIGO Scientific Collaboration}},
	date = {2015-04-09},
}

@article{tse_quantum-enhanced_2019,
	title = {Quantum-Enhanced Advanced {LIGO} Detectors in the Era of Gravitational-Wave Astronomy},
	volume = {123},
	issn = {0031-9007, 1079-7114},
	url = {https://link.aps.org/doi/10.1103/PhysRevLett.123.231107},
	doi = {10.1103/PhysRevLett.123.231107},
	pages = {231107},
	number = {23},
	journaltitle = {Physical Review Letters},
	shortjournal = {Phys. Rev. Lett.},
	author = {Tse, M. and et al.},
	date = {2019-12-05},
	langid = {english},
}

@article{acernese_increasing_2019,
	title = {Increasing the Astrophysical Reach of the Advanced Virgo Detector via the Application of Squeezed Vacuum States of Light},
	volume = {123},
	issn = {0031-9007, 1079-7114},
	url = {https://link.aps.org/doi/10.1103/PhysRevLett.123.231108},
	doi = {10.1103/PhysRevLett.123.231108},
	pages = {231108},
	number = {23},
	journaltitle = {Physical Review Letters},
	shortjournal = {Phys. Rev. Lett.},
	author = {Acernese, F. and et al.},
	date = {2019-12-05},
	langid = {english},
}

@online{et_steering_committee_et_2020,
	title = {{ET} design report update 2020},
	url = {https://apps.et-gw.eu/tds/?content=3&r=17245},
	author = {{ET steering committee}},
	date = {2020-11-29},
	note = {{ET}-00007B-20},
}

@article{ganapathy_broadband_2023,
	title = {Broadband Quantum Enhancement of the {LIGO} Detectors with Frequency-Dependent Squeezing},
	volume = {13},
	issn = {2160-3308},
	doi = {10.1103/PhysRevX.13.041021},
	pages = {041021},
	number = {4},
	journaltitle = {Physical Review X},
	shortjournal = {Phys. Rev. X},
	author = {Ganapathy, D. and et al.},
	date = {2023-10-30},
	langid = {english},
}

@article{hogeveen,
	title = {Fabry-{Perot} interferometers with three mirrors},
	volume = {25},
	copyright = {© 1986 Optical Society of America},
	issn = {2155-3165},
	doi = {10.1364/AO.25.004181},
	language = {EN},
	number = {22},
	journal = {Applied Optics},
	author = {Hogeveen, Sake J. and Stadt, Herman van de},
	month = nov,
	year = {1986},
	pages = {4181--4184},
}

@article{stadt,
	title = {Multimirror {Fabry}–{Perot} interferometers},
	volume = {2},
	copyright = {© 1985 Optical Society of America},
	issn = {1520-8532},
	doi = {10.1364/JOSAA.2.001363},
	language = {EN},
	number = {8},
	journal = {JOSA A},
	author = {Stadt, Herman van de and Muller, Johan M.},
	month = aug,
	year = {1985},
	pages = {1363--1370},
}

@article{croquette_recent_2023,
	title = {Recent advances toward mesoscopic quantum optomechanics},
	volume = {5},
	issn = {2639-0213},
	doi = {10.1116/5.0128487},
	language = {en},
	number = {1},
	journal = {AVS Quantum Science},
	author = {Croquette, M. and et al.},
	month = mar,
	year = {2023},
	pages = {014403},
}

@article{stevens_resonant_2024,
	title = {Resonant behavior and stability of a linear three-mirror cavity},
	volume = {42},
	issn = {0264-9381, 1361-6382},
	url = {https://iopscience.iop.org/article/10.1088/1361-6382/adb827},
	doi = {10.1088/1361-6382/adb827},
	pages = {065014},
	number = {6},
	journaltitle = {Classical and Quantum Gravity},
	shortjournal = {Class. Quantum Grav.},
	author = {Stevens, P. and et al.},
	date = {2025-03-21},
}

@article{strain_experimental_1994,
	title = {Experimental demonstration of the use of a Fabry–Perot cavity as a mirror of variable reflectivity},
	volume = {65},
	issn = {0034-6748, 1089-7623},
	doi = {10.1063/1.1144903},
	pages = {799--802},
	number = {4},
	journaltitle = {Review of Scientific Instruments},
	author = {Strain, K. A. and Hough, J.},
	date = {1994-04-01},
	langid = {english},
}

@article{acernese_variable_2006,
	title = {The variable finesse locking technique},
	volume = {23},
	issn = {0264-9381, 1361-6382},
	url = {https://iopscience.iop.org/article/10.1088/0264-9381/23/8/S12},
	doi = {10.1088/0264-9381/23/8/S12},
	pages = {S85--S89},
	number = {8},
	journaltitle = {Classical and Quantum Gravity},
	shortjournal = {Class. Quantum Grav.},
	author = {Acernese, F and et al.},
	urldate = {2025-05-06},
	date = {2006-04-21},
}

@article{ding_performance_2025,
	title = {Performance of multiple filter-cavity schemes for frequency-dependent squeezing in gravitational-wave detectors},
	volume = {112},
	issn = {2470-0010, 2470-0029},
	url = {https://link.aps.org/doi/10.1103/4s4d-9mb4},
	doi = {10.1103/4s4d-9mb4},
	language = {en},
	number = {12},
	journal = {Physical Review D},
	author = {Ding, Jacques and Capocasa, Eleonora and Ahrend, Isander and Liu, Fangfei and Zhao, Yuhang and Barsuglia, Matteo},
	month = dec,
	year = {2025},
	pages = {122001},
}

@thesis{stevens_study_2025,
	title = {Study of an innovative scheme for frequency-dependent squeezing in third generation gravitational-wave detectors},
	rights = {Licence Etalab},
	url = {https://theses.fr/2025UPASP174},
	institution = {Université Paris-Saclay},
	type = {PhD Thesis},
	author = {Stevens, Paul},
	editora = {Leroy, Nicolas and Lartaux-Vollard, Angélique},
	editoratype = {collaborator},
	date = {2025-12-15},
}

@misc{g_flexibility_2025,
	title = {Flexibility analysis of {ET}-{LF} Squeezing Filter Cavities {ET}-0098A-25},
	author = {Giacomo, Cianni and et al.},
	date = {2025-03-31},
}

\end{document}